\documentclass[a4paper,11pt]{article}
\usepackage{jinstpub} 
\usepackage{lineno}
\usepackage{siunitx}
\usepackage{subcaption}

\title{\boldmath Design and Development of Portable RPC-Based Cosmic Muon Tracker}

\author[a,1]{Yuvaraj Elangovan,\note{Corresponding author.}}
\author[a,2]{B. Satyanarayana,} 
\author[a]{Ravindra Shinde,} 
\author[a]{Mandar Saraf,}
\author[a]{Pathaleswar,}
\author[a]{S. Thoi Thoi,} 
 \author[a]{Gobinda Majumder,} 
\author[a]{S.R Joshi,} 
\author[a]{Piyush Verma,}
\author[b]{Honey Khindri,} 
\author[a]{and Umesh L} 
\affiliation[a]{Tata Institute of Fundamental Research, Mumbai, India}
\affiliation[b]{Homi Bhabha National Institute, Mumbai, India}

\emailAdd{yue8@pitt.edu}

\abstract{
Primary cosmic rays when interact with the atmosphere, produce a cascade of lighter secondary particles namely pion, kaon, neutrons, muons, electrons, positrons and neutrinos. Muons are one of the most abundant and easily detectable particles at the ground surface using a large variety of particle detectors. Resistive Plate Chambers (RPCs) of $2\, \text{m} \times 2\, \text{m}$ in dimension were developed to be used in large scale as the active detector elements in the Iron Calorimeter (ICAL) which was planned to be built by the India-based Neutrino Observatory (INO). As a spin-off of this work, a portable stack of eight 26~cm~\(\times\)~26~cm RPC detectors was developed, named the Cosmic Muon Tracker (CMT). It could be used to conduct small-scale particle detector experiments as well as training Students. We will discuss design, integration, characterization and some of the applications of this detector in this paper.
}


\begin{document}
\maketitle
\flushbottom

\section{Introduction}
\label{sec:intro}
High-energy cosmic rays, primarily protons, interact with Earth's atmosphere, produce showers of secondary particles mostly pions. These pions further decay into muons and neutrinos. This shower of particles initially increases and then reduces in number as they travel towards the earth's surface, resulting in a measurable muon flux of approximately one muon per square centimeter per minute at sea level.  Muon flux varies with altitude \cite{muon_flux} and their angular distribution is influenced by the Earth's magnetic field at a given location. Due to their wide spectrum of energy, angular distributions and abundance, muons are used as radiation probe in diverse fields, including medical imaging, geophysical tomography, and high-energy physics. Several detector technologies have been used for muon detection, such as plastic scintillators~\cite{plastic}, liquid scintillators~\cite{liquid}, time projection chambers (TPCs)~\cite{tpc}, semiconductor detectors detectors~\cite{silicon} and Resistive Plate Chambers (RPCs)~\cite{rpc}.  Among them, RPC detectors offer advantages such as excellent time resolution, large-area coverage, ease of fabrication and cost-effectiveness. In general, multiple large area RPCs arranged in a stacked geometry are used to track muons. 

\begin{figure}[htbp]
\centering
\includegraphics[width=.5\textwidth]{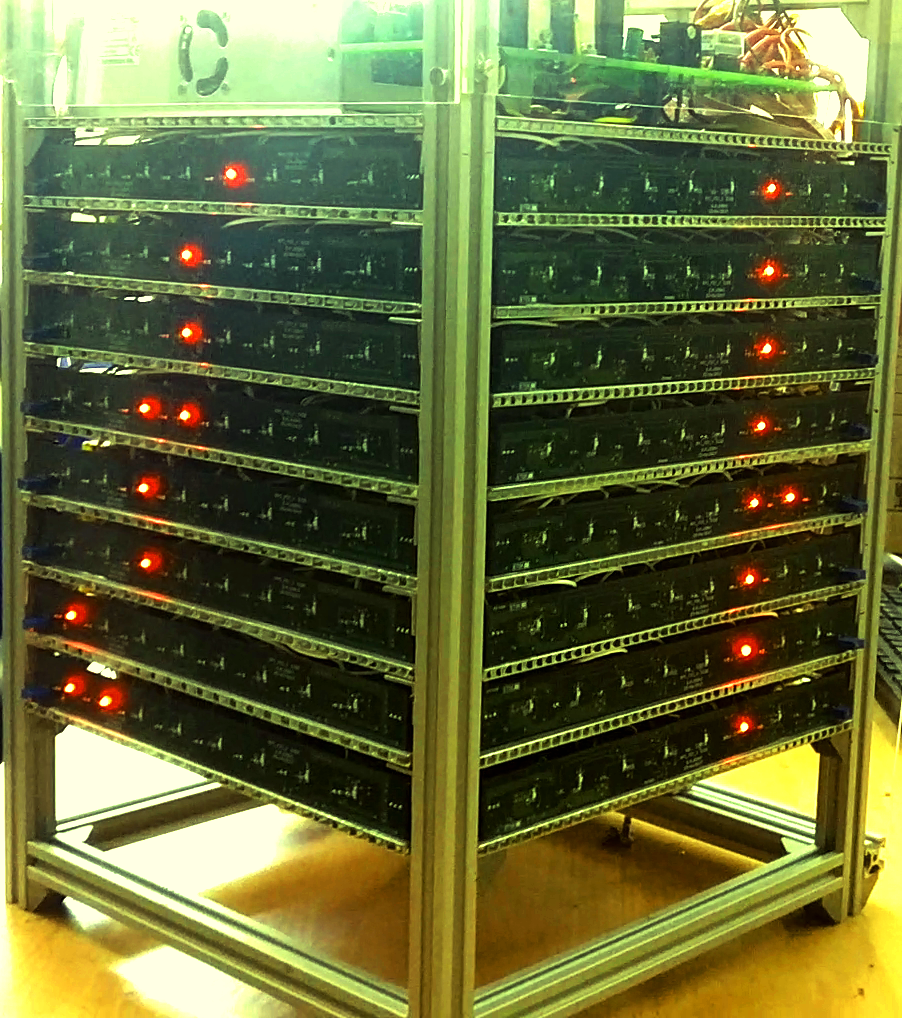}
\caption{Photograph of the CMT displaying a muon track. The track is visualized in real time using LEDs mounted on the front-end boards connected to the RPC detector stack\label{fig:1}}
\end{figure}

The Iron CALorimeter (ICAL) experiment, proposed by the India-based Neutrino Observatory (INO) collaboration \cite{ino_report}, aims to track and study atmospheric neutrinos using a 50-kiloton iron calorimeter. The calorimeter comprises of 28,800 RPCs, each measuring \(2~\mathrm{m} \times 2~\mathrm{m}\), sandwiched between 56 \text{mm} thick iron plates to track charged particles produced by neutrino interactions. Each RPC is equipped with a Front-End Data Acquisition module, known as RPC-DAQ, which records the position and timing information of particle trajectories. The collected data is transmitted to back-end servers via an Ethernet interface, with each RPC-DAQ remotely controllable through a dedicated IP address. Event registration and synchronization are managed using a Real-Time Clock system. Based on the detector and instrumentation R\&D for the ICAL experiment, we developed a compact, eight-layer RPC-based detector stack—named the Cosmic Muon Tracker (CMT)—that demonstrates real-time muon tracking and serves as a valuable tool for education and outreach. The detector stack consists of eight RPCs, each measuring approximately \(26~\text{cm} \times 26~\text{cm}\). These detectors are filled with a gas mixture and tightly packed for operational stability. Each RPC contains eight equally segmented signal strips on both the top and bottom surfaces.   These strips are placed orthogonally and marked as X- and Y-strips, respectively, to enable two-dimensional position readout. Cosmic ray muons typically traverse multiple RPCs in the stack, inducing signals on the pickup strips intersected by their trajectories. LEDs—one per strip—mounted on the front-end boards of the X- and Y-planes illuminate in real time to display the muon tracks, as shown in Figure~\ref{fig:1}. To enable real-time visualization and recording of muon events, the CMT uses a compact FPGA-based Data Acquisition (DAQ) system. This system captures signals from all RPC layers and uses configurable logic to identify valid muon events. Upon detection, the corresponding strip hits are both logged and displayed using on-board LED matrices. The compact design, along with locally generated low and high voltage supplies from standard AC mains, makes the entire setup self-contained and highly portable. A more detailed description of the trigger logic and DAQ architecture is provided in Section~\ref{sec:daq}. The paper describes design and fabrication of the detectors, electronics, trigger and data acquisition systems, performance of the stack as well as its other potential applications.

\section{RPC Design Overview and Detection Principle}
\label{sec:principle}
\begin{figure}[htbp]
\centering
\includegraphics[width=.9\textwidth]{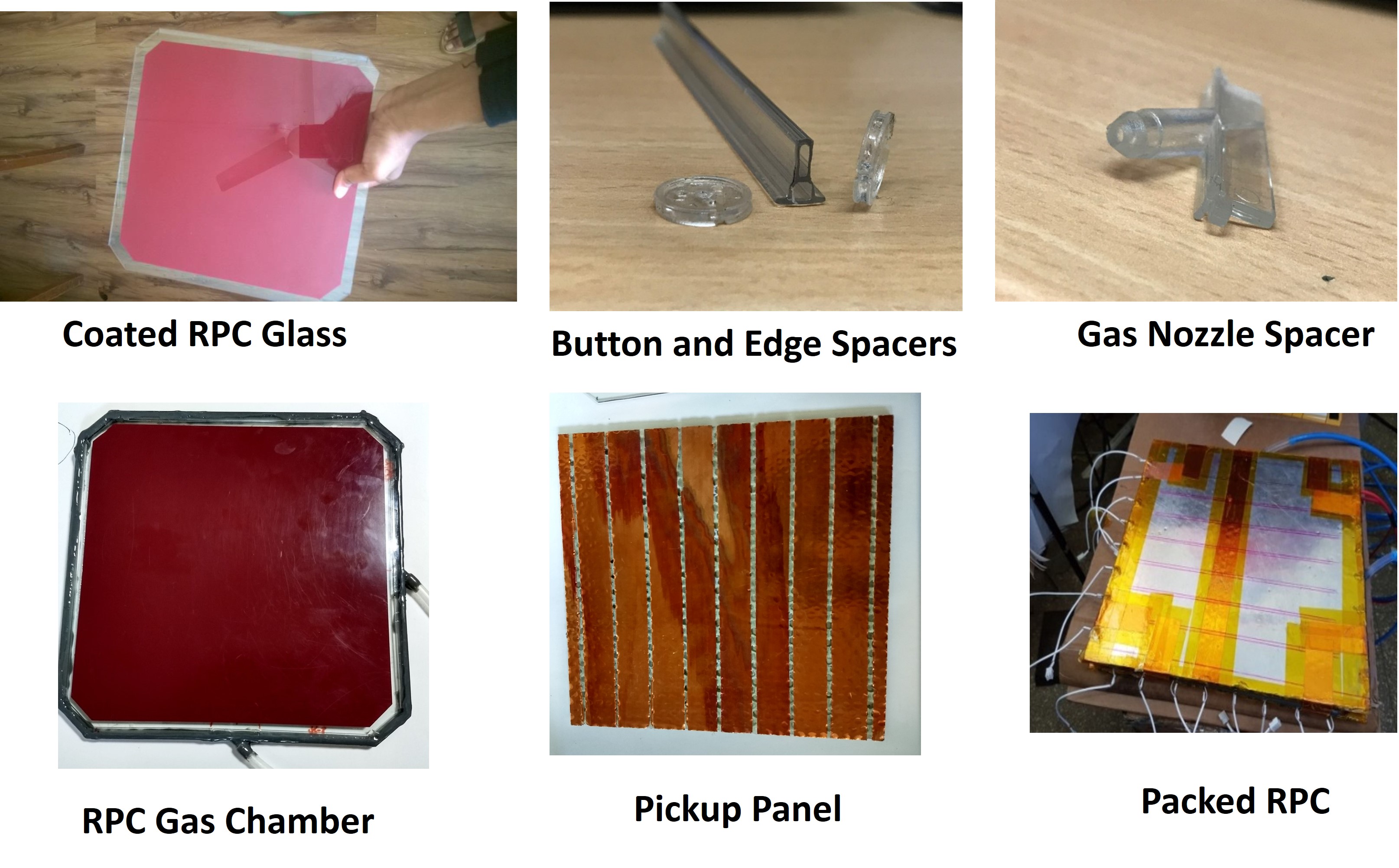}
\caption{Major components involved in the fabrication of a single RPC detector. The images show (from left to right, top to bottom): Coated glass sheet, edge and button spacers, gas nozzles, a fully sealed gas chamber, pickup panels with copper strips, and a completed RPC unit. \label{fig:2}}
\end{figure}
Each RPC comprises of two parallel 2~mm-thick glass plates with a bulk resistivity of approximately \(10^{12}~\Omega\cdot\mathrm{cm}\) as measured at room temperature \cite{resistivity}. A semi-conductive paint is coated on one side of the glass plates. These plates are mounted parallel to each other - with uncoated sides facing each other. 2~mm thick polycarbonate button-shaped spacers are placed uniformly within the gap to maintain the separation. Edge and nozzle spacers are used to maintain uniform separation along the chamber edges, ensuring both mechanical stability and proper gas flow throughout the RPC volume. List of major components used to fabricate RPC is shown in Figure~\ref{fig:2}. 3M DP-190 epoxy adhesive was used to seal the chamber perimeter after assembling the plates and spacers. DP-190 was chosen for the INO ICAL RPCs for its non-reactive properties with most of the detector gases. Epoxy adhesive may exhibit mild out gassing, but no degradation in RPC performance (e.g., leakage current or efficiency) was observed during prior studies \cite{rpc}. A bias voltage of about $10\, \text{kV}$ is applied across the outer coated surfaces, establishing a strong and uniform electric field inside the gap. 
The RPCs operate in avalanche mode with a gas mixture of R-134a (95.0\%), iso-butane (4.2\%), and SF6 (0.3\%). R-134a serves as the primary ionizing medium, providing a high number of electron-ion pairs, iso-butane acts as a photon quencher to suppress secondary avalanches and stabilize operation, and SF6 functions as an electron quencher to minimize streamer formation. This composition is widely used in several large-scale RPC-based experiments \cite{gas} and has been time-tested in the long-term operation of INO-ICAL RPCs.

\begin{figure}[htbp]
\centering
\includegraphics[width=.85\textwidth]{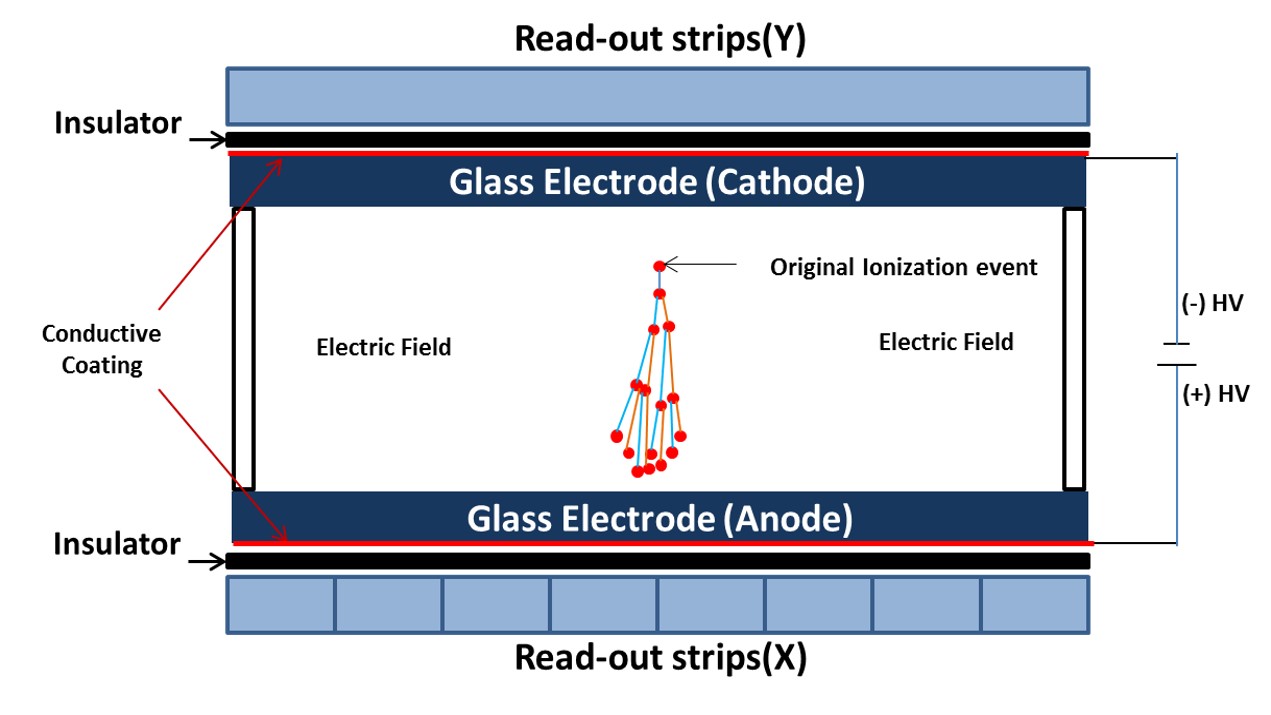}
\caption{Schematic side view illustrating the ionization and avalanche process inside a RPC. \label{fig:3}}
\end{figure}

When a charged particle, such as a cosmic ray muon, passes through the gas-filled RPC, it can ionize gas molecules along its path, producing free electrons as well as positive and negative ions. The electrons are accelerated towards the anode by due to the electric field, generating a series of secondary ionizations and leading to an electron avalanche shown in Figure~\ref{fig:3}.
This avalanche induces an electric signal on the copper strips. Each side of the RPC is divided into eight copper strips, each of width $28\, \text{mm}$ and a pitch of $30\, \text{mm}$. The resistive nature of the glass plates ensures that the discharge is localized, and prevents it from spreading across the entire plate. Thin Mylar sheets are used to isolate the pickup strips from the semi-conductive coating carrying high voltage. The pickup strips are appropriately terminated to readout the signal from one end only. The induced signals are transmitted to the front-end electronics for amplification, discrimination, and readout.

\section{RPC Production and Testing}
\label{sec:production}
The RPCs used in the CMT were fabricated and tested in-house through a well-defined quality control process \cite{qc}. Undergraduate and graduate students from various institutions actively participated in the fabrication, quality checks, and characterization of RPCs as part of summer training and laboratory programs, gaining hands-on experience in detector construction. Students assisted with various aspects of the assembly and testing workflow. These included surface resistivity measurements of the resistive coating, leak testing of the gas gap, current-voltage (I--V) characterization, and efficiency measurements using cosmic ray muons. The testing process followed standardized protocols developed for large-scale RPC deployment in the INO-ICAL experiment. Details of the individual test procedures and results are presented in the following subsections.

\subsection{Surface Resistivity Measurement}
\label{sec:res}

\begin{figure}[htbp]
\centering
\includegraphics[width=.9\textwidth]{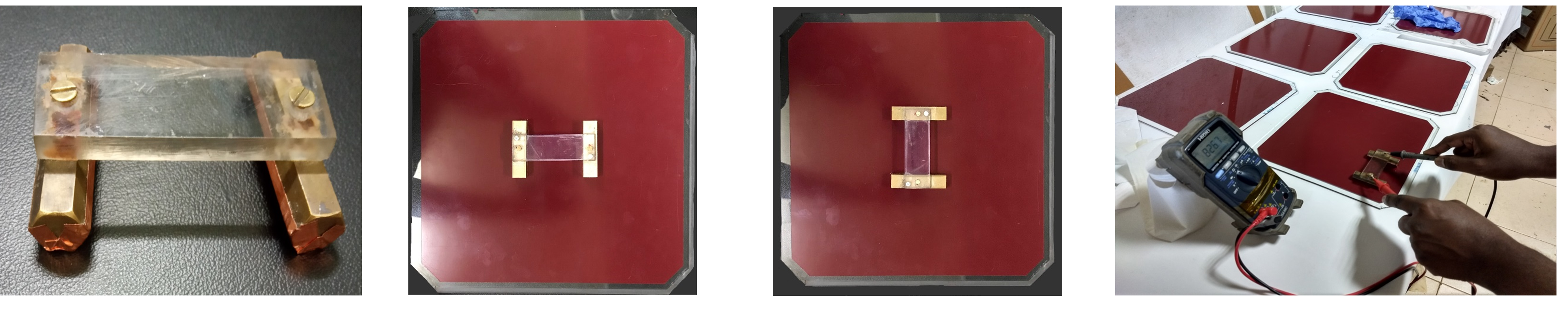}
\caption{Setup for surface resistivity measurement: Resistivity measurement jig, Placement in two orthogonal orientations (“H” and “I”) and Probing process using a digital multimeter.\label{fig:4}}
\end{figure}

Uniform surface resistivity of the resistive paint coating on the outer glass surfaces is essential to ensure uniform voltage distribution and electric field uniformity across the RPC \cite{sur_res}. To measure this, a custom-built jig was used in conjunction with a digital multimeter. The jig consists of two parallel brass electrodes, separated by an insulating spacer, forming a well-defined square contact area on the painted surface. Measurements were performed on a $3 \times 3$ grid covering each coated glass plate. The jig was placed in two orthogonal orientations as shown in Figure~\ref{fig:4}, labeled ``H'' and ``I'', to account for any directional variation in coating uniformity. 

\begin{figure}[htbp]
\centering
\includegraphics[width=.7\textwidth]{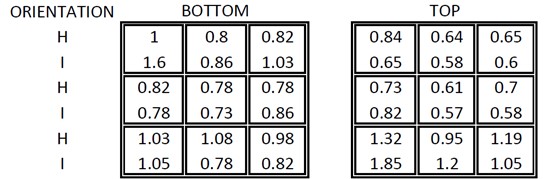}
\caption{Measured surface resistivity values (in M$\Omega$) for a single RPC glass plate in both “H” and “I” jig orientations.\label{fig:5}}
\end{figure}

Surface resistivity measurements across the glass surface were generally centered near \(1~\mathrm{M\Omega}\), with observed variations up to a factor of 3 in some cases, as shown in Figure~\ref{fig:5}. These values remain within acceptable limits for stable RPC performance. It may be noted that the quantity measured by the multi-meter is resistance, while it actually represents a quantity called “resistance per square” due to construction of the jig as described above. These values are consistent with operational requirements for RPCs and comparable to those used in the INO-ICAL prototypes.

\subsection{Leak test of RPC Gaps}
\label{sec:leak}
The fabricated gas gaps must be gas-tight else the performance of the RPC could be compromised due to the contaminated gas mixture. To ensure that the RPCs are leak free, leak test measurements \cite{leak} were conducted using an Arduino-based system. The outlet of the RPC was connected to a pressure sensor linked to a NodeMCU microcontroller Wi-Fi module which transmits the pressure readings inside the chamber directly to a server. The inlet of the RPC was connected to a gas supply with a needle valve to finely adjust the pressure filling the chamber with nitrogen gas. Another NodeMCU based pressure sensor module was used to measure the atmospheric pressure readings and transmitted to the server. A Python script was used to store the data and display the pressure difference over time. The pressure inside the RPC was increased above the atmospheric pressure and tightly sealed. The pressure inside the RPC and pressure difference are continuously monitored and logged at least for 5 hours. If the pressure remained stable within its initial range, then the chamber was deemed to be leak-free and ready for further fabrication. The temperature inside the chamber was also plotted as a function of time. 

\begin{figure}[htbp]
\centering
\includegraphics[width=.8\textwidth]{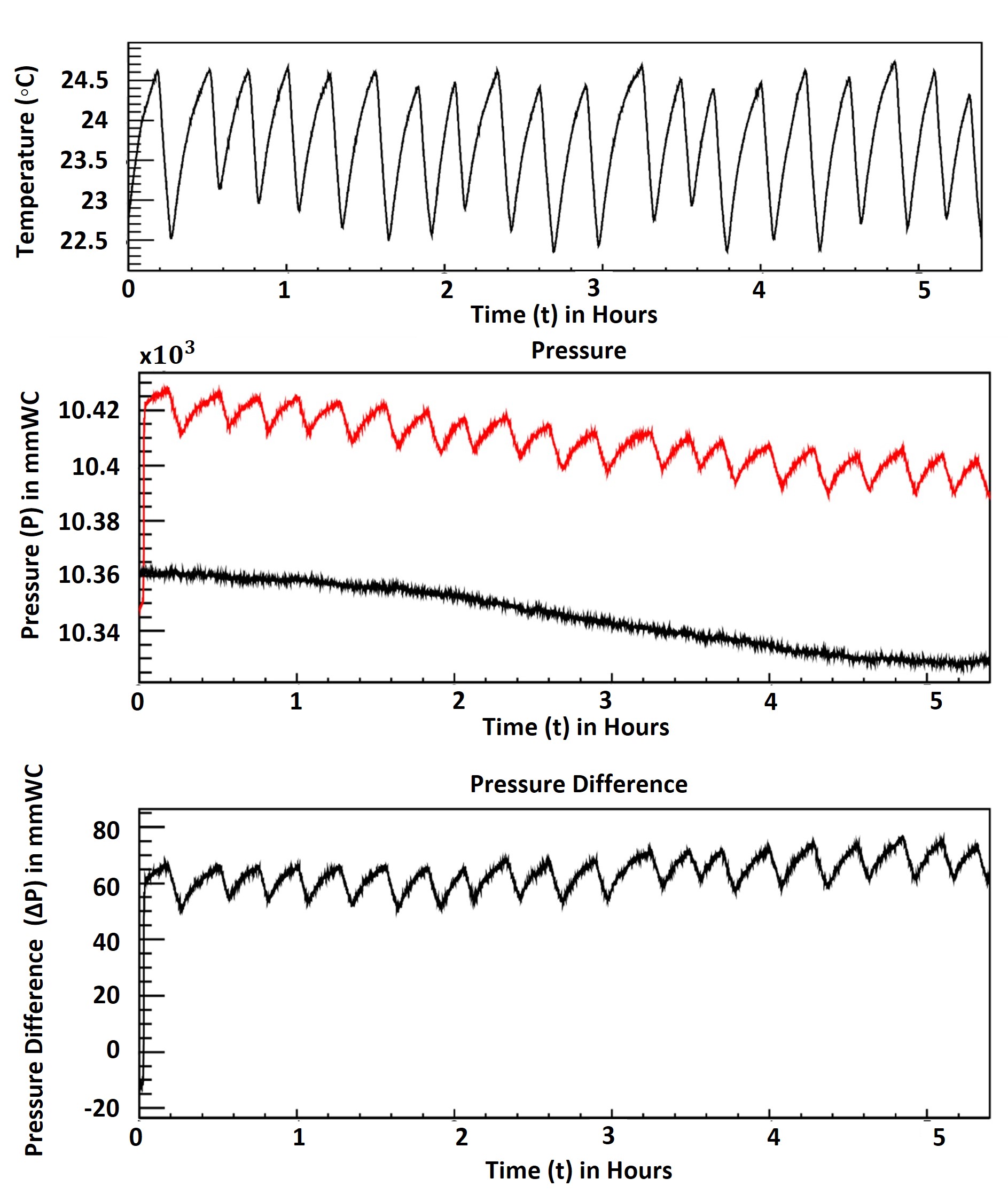}
\caption{Leak test results for a sealed RPC chamber. The top panel shows laboratory temperature as a function of time. The middle panel displays the internal pressure of the RPC (red) and atmospheric pressure (black). The bottom panel shows the difference in pressure between the RPC and ambient atmosphere over time.\label{fig:6}}
\end{figure}

A temperature variation of 22.5$^\circ$C. to 24.5$^\circ$C was observed due to the ON-OFF duty cycle of the lab air conditioners. In case a gas leak was found, a hand-held gas leak tester was used to locate the exact position of the leak and the RPC was re-glued. Typical leak test plots include temperature variation (mostly due to air conditioning), atmospheric pressure (black), pressure inside the chamber (red), and the pressure difference as shown in the bottom plot of Figure~\ref{fig:6}. The leak-test data similar to that of the electrode paint resistance measurements was documented for all RPCs for future reference.


\subsection{I-V Characterization}
\label{sec:iv}
\begin{figure}[htbp]
\centering
\includegraphics[width=.35\textwidth]{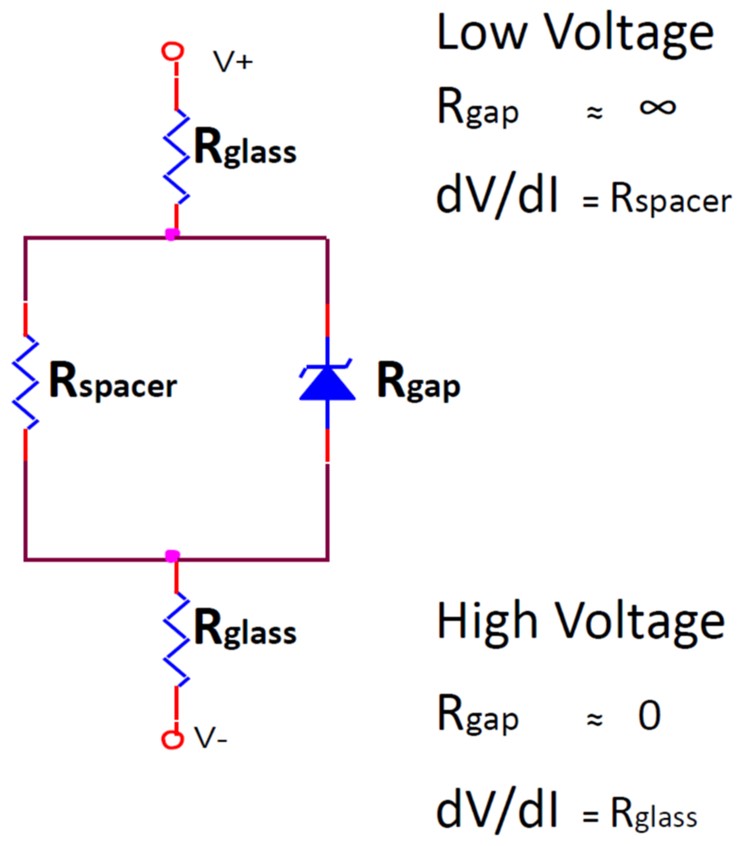}
\caption{Equivalent Circuit of an RPC.\label{fig:7}}
\end{figure}

Figure~\ref{fig:7} shows the equivalent circuit of an RPC gas gap, where the leakage current at low bias is primarily attributed to conduction through the polycarbonate spacers, denoted as resistance \( R_{\text{spacer}} \) between the plates. Epoxy used for sealing the chambers contributes negligibly to the overall current. No signs of surface leakage across epoxy-glass boundaries were observed during testing, as confirmed by stable current readings and visual inspection of glue lines. When the RPC attains the operating potential, an electric field is established between the top and bottom plates and the current begins to flow through the gas medium whose resistance is represented as $R_{gap}$.

For the I-V characteristics test, the total high voltage applied to the gap was increased from zero to 10000 kV in steps of 1000 V. Channel A corresponds to positive polarity and channel B to negative polarity, meaning that a total potential difference of $10000\, \text{V}$ is applied when both channels reach a voltage of $5000\, \text{V}$ each. Current readings were recorded at each step after $15\, \text{minutes}$ allowing the currents to stabilize. 

\begin{figure}[htbp]
\centering
\includegraphics[width=0.9\textwidth]{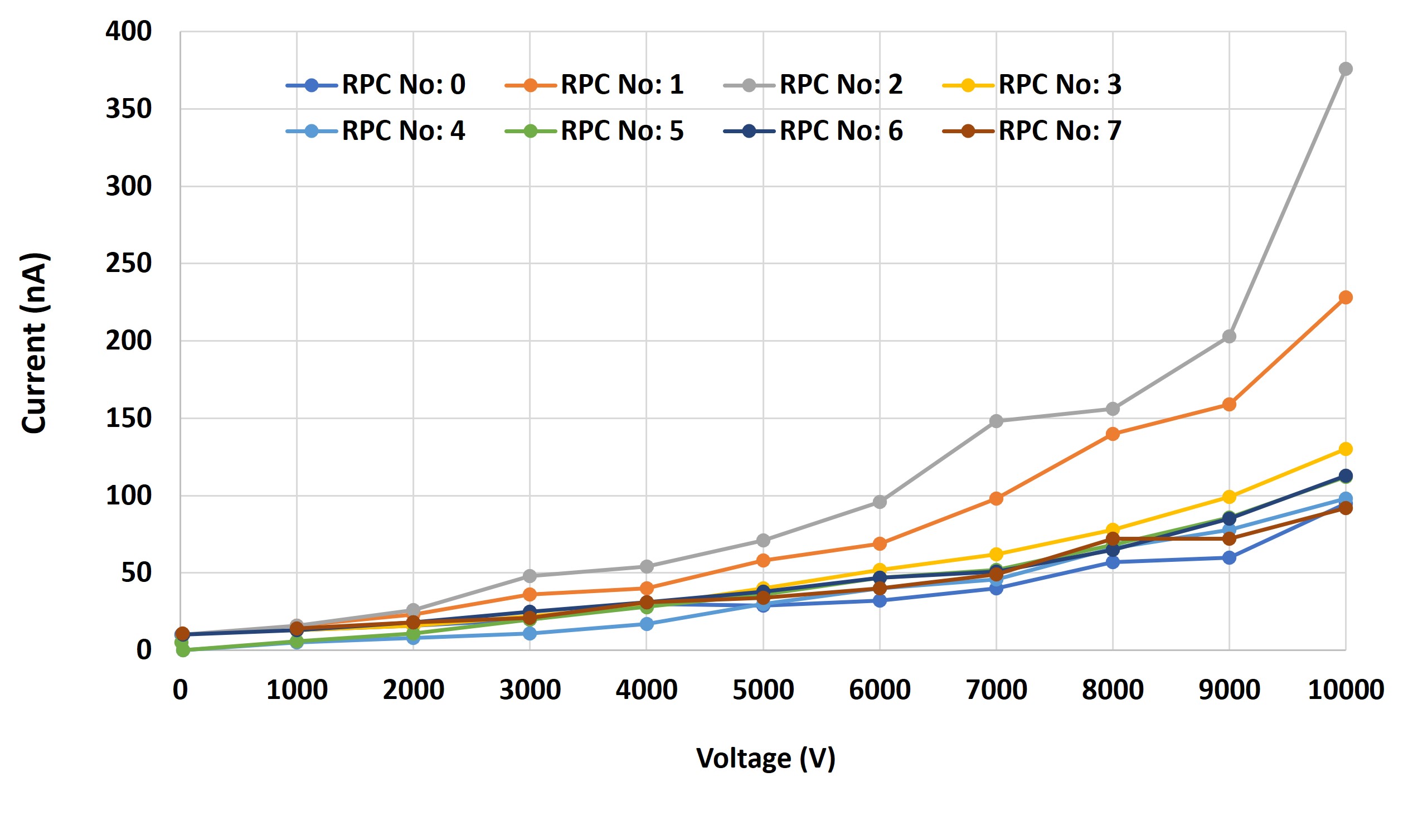}
\caption{I-V Characteristics of 8 RPCs.\label{fig:8}}
\end{figure}

Figure~\ref{fig:8} shows the current–voltage (IV) characteristics for all eight RPCs. While most chambers exhibit similar current behavior, RPC 1 and RPC 2 show slightly elevated currents across the bias range. The deviation in current observed at low voltages (2000–3000\,V), where gas gain is negligible, is likely due to leakage currents across the spacers and the chamber borders. While surface irregularities or minor contamination during assembly may amplify these effects.
However, these values remained well within safe operating limits and showed no sign of electrical breakdown or instability during extended testing.

\subsection{RPC Detector Characterization}
\label{sec:char}
Efficiency of the RPCs are measured using a set of reference scintillator paddles which are optically coupled to a photomultiplier tube (PMT). The scintillator paddles receive high voltage from the HV distribution box while the RPC is supplied with bias supply separately. The signal outputs from the scintillator paddles are fed to a discriminator module where adjustments to the threshold and pulse width can be made. Since RPC pulses are small - typically in the millivolt range, they are amplified using a Hybrid Micro Circuits (HMC) preamplifier, a transistor-based voltage amplifier with a gain of about 80 before feeding them into the discriminator modules.

\begin{figure}[htbp]
\centering
\includegraphics[width=.8\textwidth]{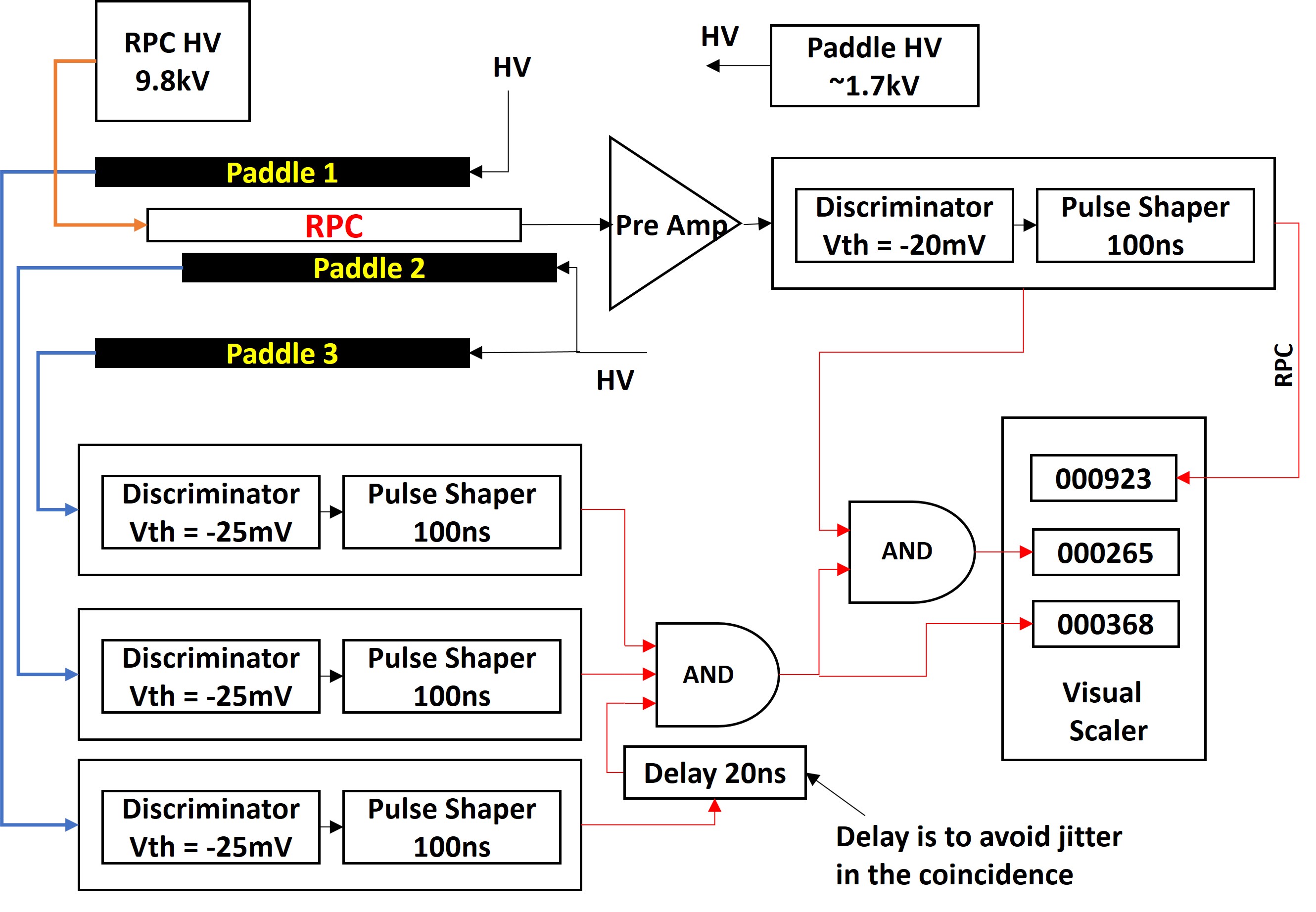}
\caption{Block diagram of the setup used for RPC characterization with plastic scintillator paddles.\label{fig:9}}
\end{figure}

\begin{figure}[htbp]
\centering
\includegraphics[width=.7\textwidth]{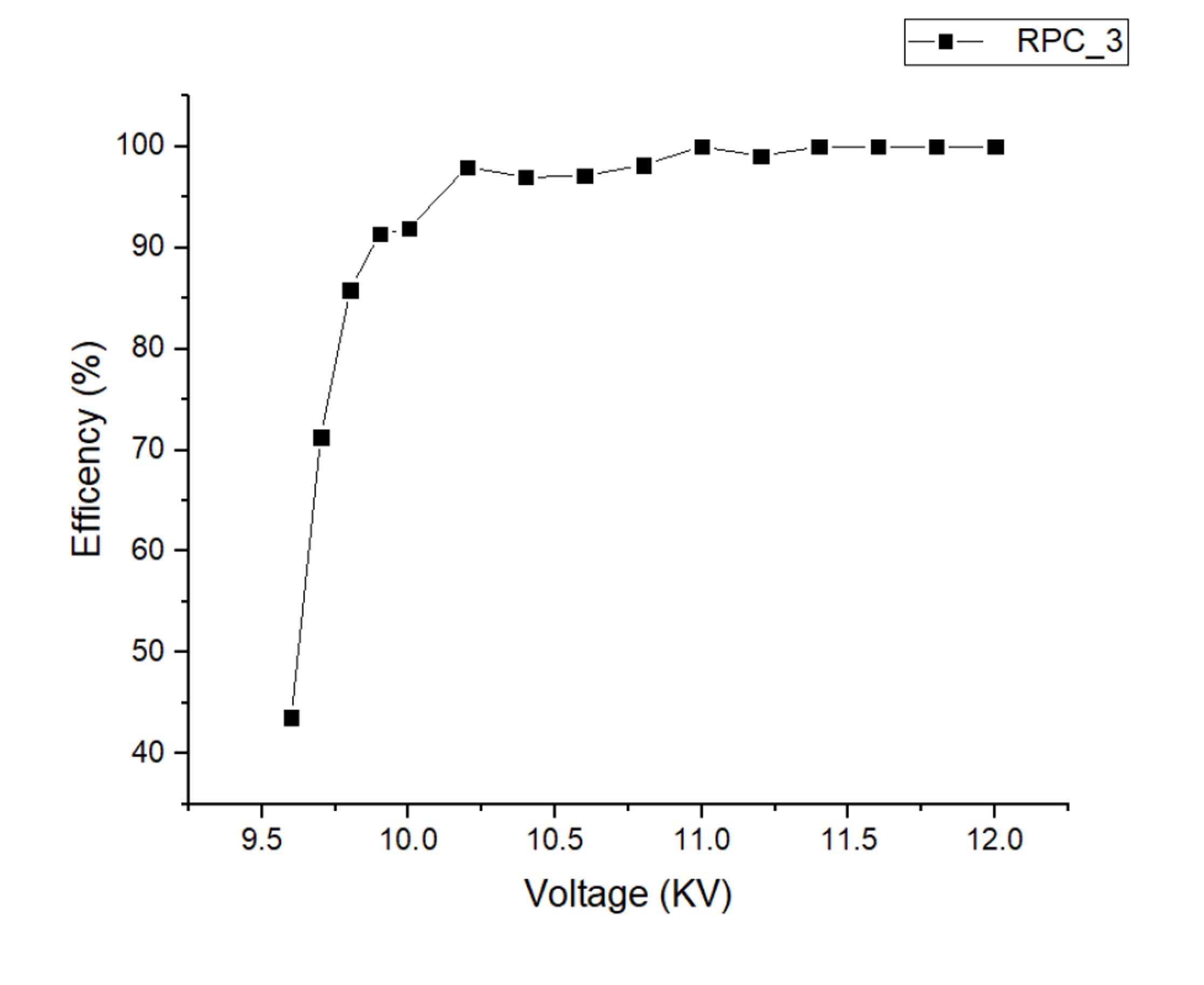}
\caption{Detection efficiency of an RPC as a function of total applied high voltage.\label{fig:10}}
\end{figure}

Three scintillator trigger paddles are aligned on top and bottom of the RPC covering a specific area of the RPC strip.  The overall test setup is shown in Figure~\ref{fig:9}. The successful muon events generate signals in all the three paddles and RPC strips. The efficiency of the RPC is determined by the ratio of the 4Fold (P1 AND P2 AND P3 AND RPC strip) to the 3Fold (P1 AND P2 AND P3). The HV bias of the RPC is slowly increased in steps of $100\, \text{V}$ varying from $9\, \text{kV}$ to $12\, \text{kV}$. At each SET voltage, the counting rates of the RPC strip signal, 3Fold and 4Fold are measured using a visual pre-scaler. The efficiency of an RPC is plotted as shown in Figure~\ref{fig:10}.

\begin{figure}[htbp]
\centering
\includegraphics[width=.9\textwidth]{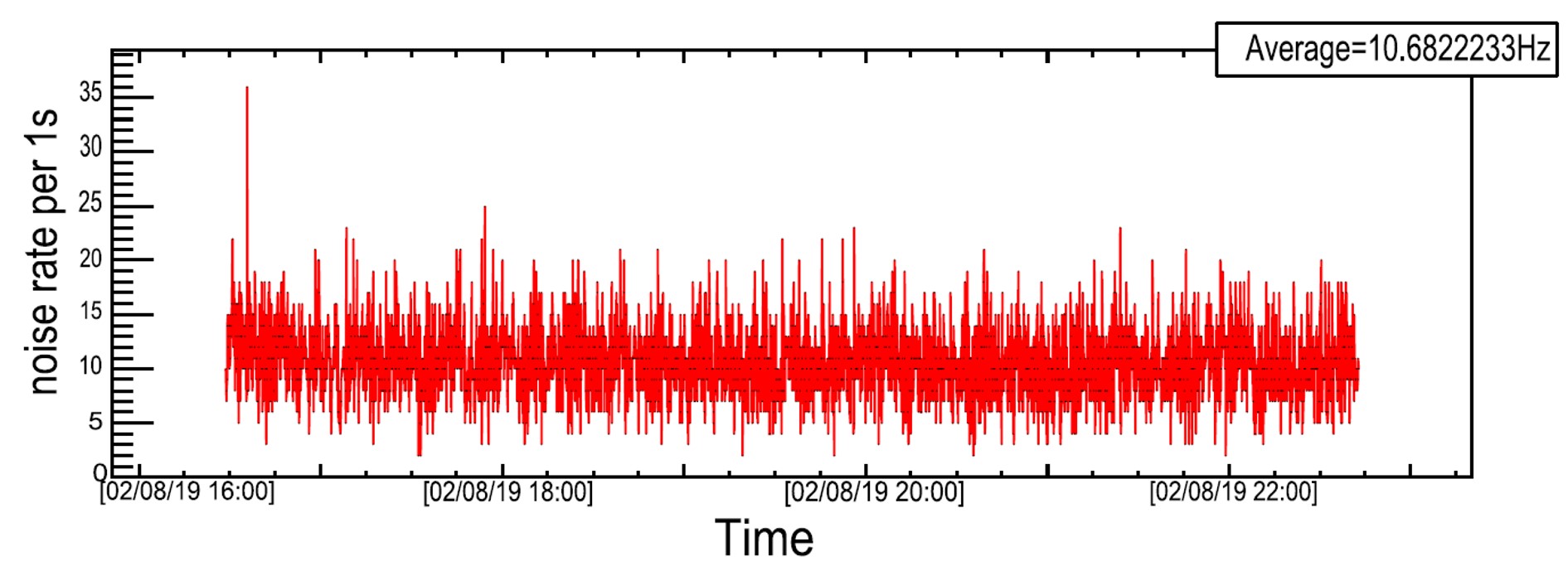}
\caption{Noise rate measurement for a single RPC channel over time.\label{fig:11}}
\end{figure}

Noise rate is one of the basic health monitoring parameters of the RPC. Noise rate of an RPC must be stable for proper tracking of muons through the RPC stack. Figure~\ref{fig:11} shows the noise rate of one of the RPC channels. The observed strip noise rates, typically in the range of 5–15 Hz, are consistent with expected values for similarly sized glass RPCs operated in avalanche mode. Similar noise distributions were observed across all eight chambers, with only a few strips requiring masking due to persistent high rates.

\section{RPC integration into  CMT Stack}
\label{sec:geo}

\begin{figure}[htbp]
\centering
\includegraphics[width=.8\textwidth]{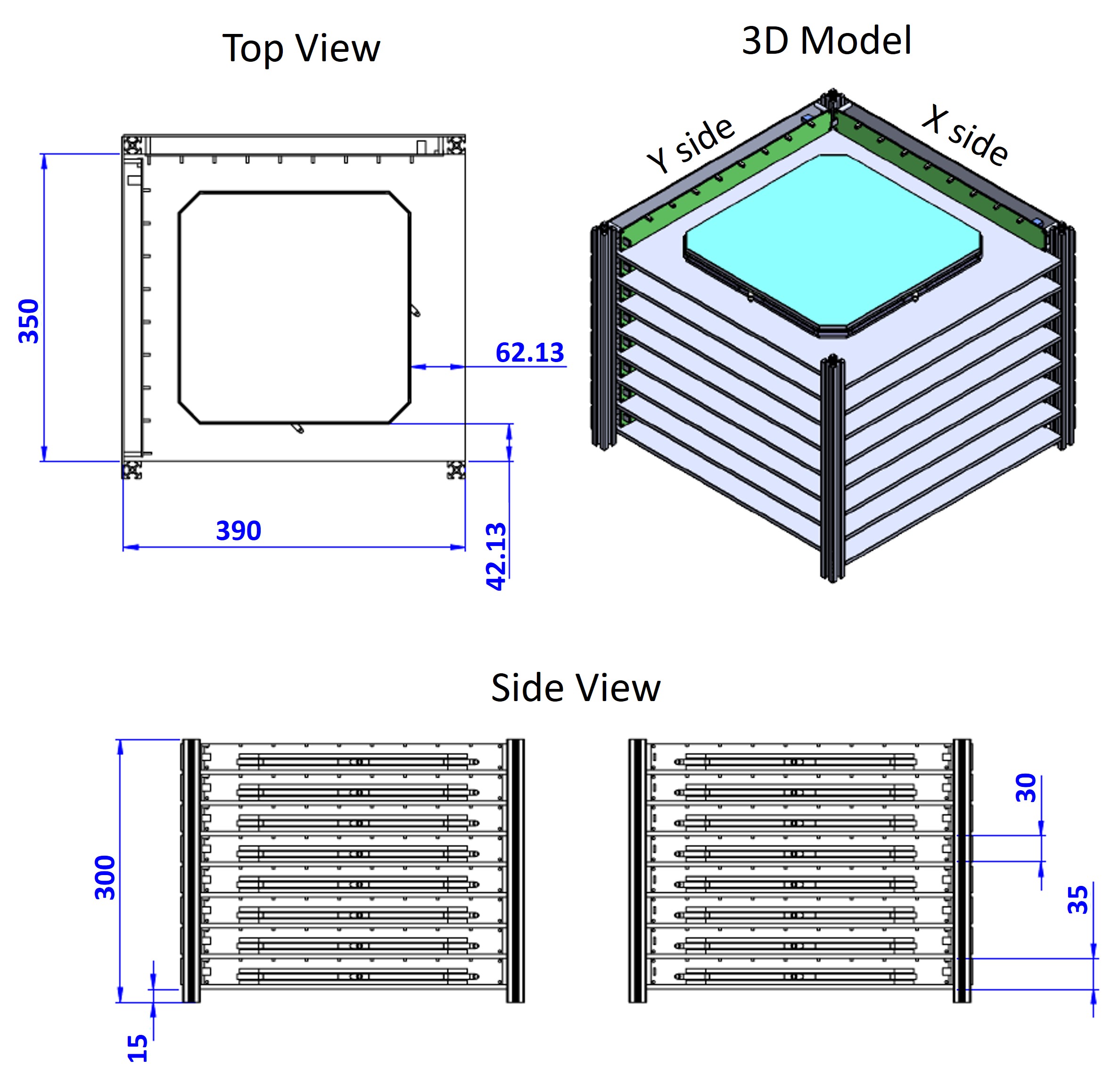}
\caption{Top view, side view, and 3D model of the 8-layer RPC detector stack.\label{fig:12}}
\end{figure}

The CMT consists of eight RPCs arranged vertically in a stack, with a uniform spacing of 35~mm between adjacent layers. Each RPC measures $26 \times 26$ $~\mathrm{cm}^2$ = $676~\mathrm{cm}^2$, with the central $24 \times 24$ $~\mathrm{cm}^2$ = $576~\mathrm{cm}^2$ region defined as the active detection area. Figure~\ref{fig:12} shows multiple views, including top, side, and a 3D rendering, of the complete eight-layer CMT assembly.

\begin{figure}[htbp]
\centering
\includegraphics[width=.6\textwidth]{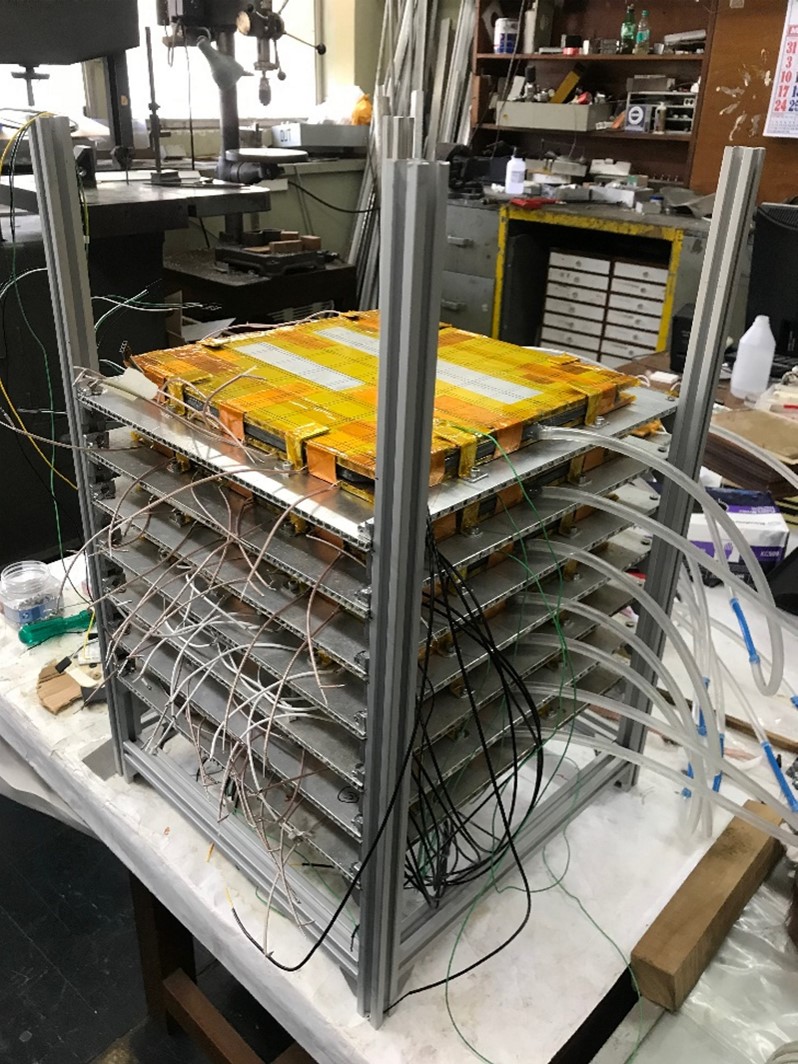}
\caption{Stacking of RPC detectors into the 8 layer frame structure.\label{fig:13}}
\end{figure}

After individual testing and characterization, Each RPC is mounted on a 3~mm-thick aluminum honeycomb base-plate using custom-fabricated Z-shaped aluminum clamps, providing both structural rigidity and weight reduction. The detectors along with base-plates are placed within a custom-designed aluminum frame shown in Figure~\ref{fig:13} using L-brackets and C-type vertical channels that provide mechanical alignment and support. The complete CMT stack houses all essential subsystems, including the RPCs, front-end electronics, power supply modules, data acquisition system, and gas manifolds. The overall size of the stack is $350\, \text{mm} \times 390\, \text{mm} \times 600\, \text{mm}$. 

Dedicated aluminum brackets were fabricated to mount the front-end boards on both X- and Y-sides of each layer. Power distribution cables for the front-end electronics were routed through designated slots in the frame.  All components were connected to a common ground point using copper strips and grounding wires. Front-end electronics are mounted on the X- and Y-sides of the stack, while high-voltage cabling and gas manifolds are arranged along the rear and top. Power and signal cables are neatly routed through the vertical frame channels, and transparent acrylic panels enclose the sides of the stack to protect the RPCs and electronics while allowing visual access. Gas lines were connected in a cascading configuration to ensure uniform gas flow through each chamber. The internal pressure remained stable over time due to the high-quality sealing of inlet and outlet fittings. High voltage for RPC operation is generated locally using a compact onboard power module, allowing the entire tracker to operate directly from a standard power source without requiring external infrastructure.

\section{Trigger and Data Acquisition System}
\label{sec:daq}
\begin{figure}[htbp]
\centering
\includegraphics[width=.7\textwidth]{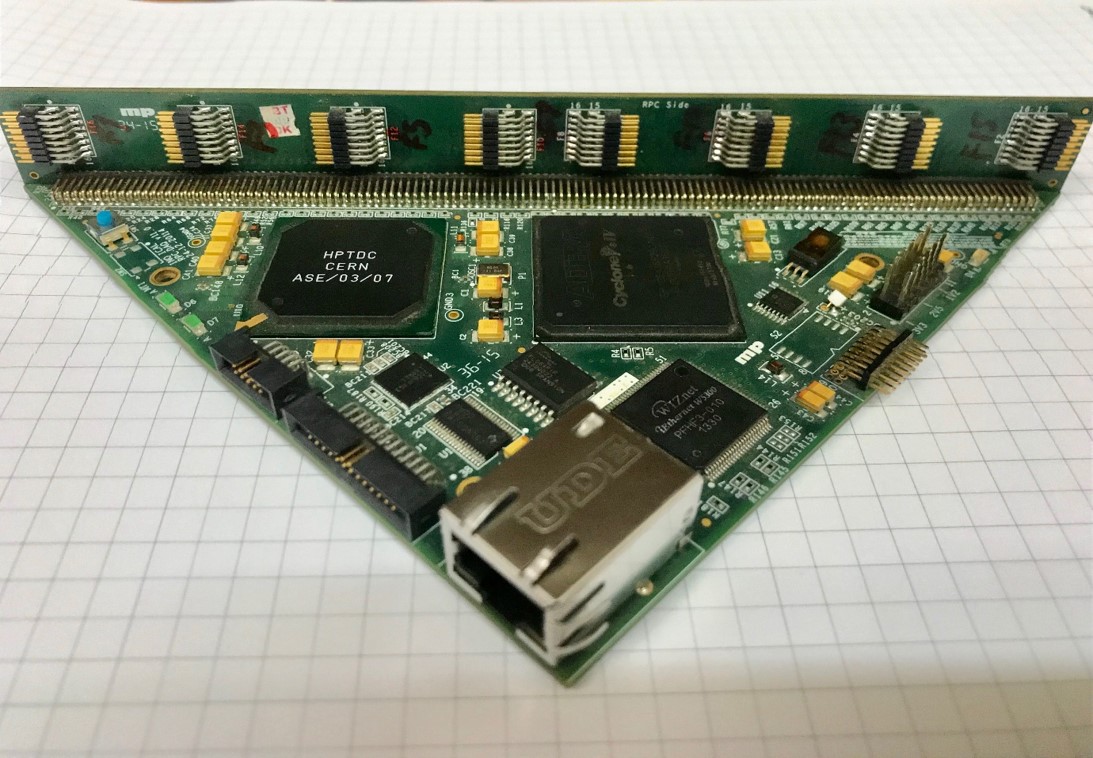}
\caption{RPC-DAQ Module developed for INO ICAL Experiment.\label{fig:14}}
\end{figure}

\begin{figure}[htbp]
\centering
\includegraphics[width=.7\textwidth]{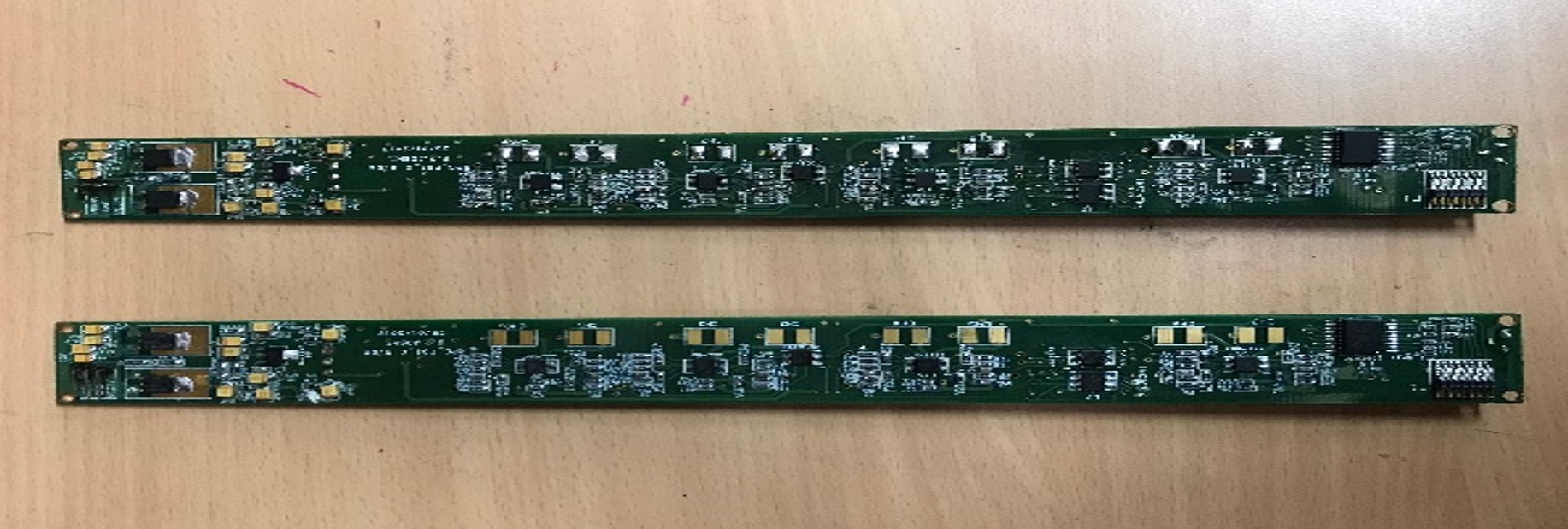}
\caption{Front-End Boards for X- and Y-side readout.\label{fig:15}}
\end{figure}

\begin{figure}[htbp]
\centering
\includegraphics[width=1\textwidth]{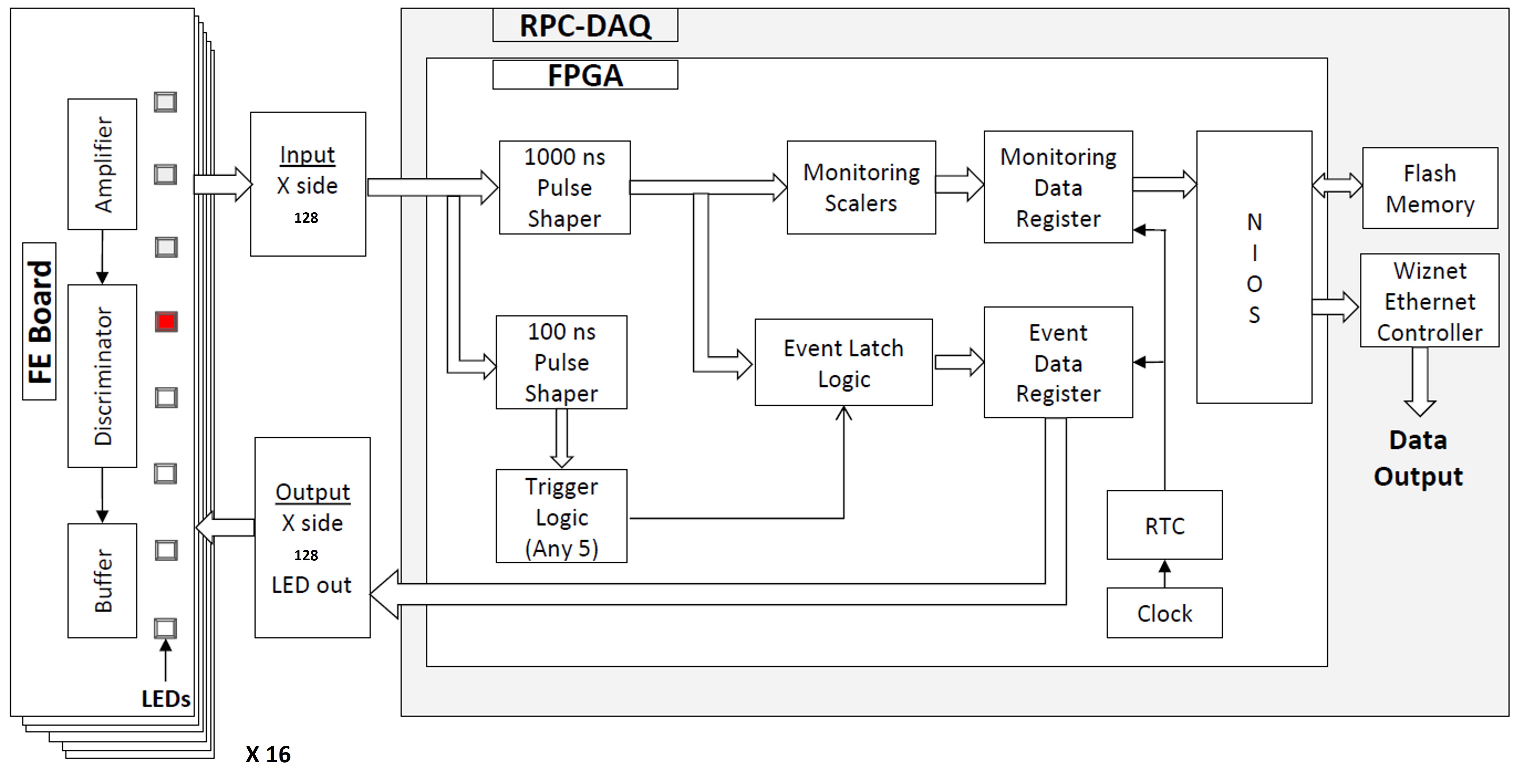}
\caption{Block diagram of the front-end readout and DAQ logic implemented inside the RPC-DAQ FPGA. \label{fig:16}}
\end{figure}

The CMT employs a custom-designed, FPGA-based DAQ module named RPC-DAQ ~\cite{rpc-daq} originally developed for the INO-ICAL experiment shown in Figure~\ref{fig:14}. RPC-DAQ is capable of handling 128 inputs covering 64 channels from X-side and 64 channels from Y-side, interfaces with 16 front-end (FE) boards shown in Figure~\ref{fig:15}: 8 for the X-plane strips and 8 for the Y-plane strips.
When a charged particle traverses the CMT, it induces coincident signals in aligned strips of multiple RPC layers. Each FE board amplifies 8 analog inputs and outputs discriminated digital signals via Low Voltage Differential Signaling  (LVDS) lines to the DAQ module. The DAQ logic shown in Figure~\ref{fig:16} implements configurable trigger conditions based on coincidence between user-defined layers. When a valid trigger is detected - indicating passage of a cosmic muon through the stack, the hit pattern is latched and stored.

Simultaneously, the corresponding strip-level LEDs are activated, producing a real-time visual display of the muon track across the stack. This latched state is maintained until the next trigger event. The digitized hit data is timestamped using a High-Performance Time-to-Digital Converter (HPTDC) \cite{hptdc} with $100\, \text{ps}$ resolution, This helps in tracking muons with accurate directional measurement. Also, a $100\, \text{ns}$ Real-Time Clock (RTC) is used to timestamp every muon event. An Event data comprises of position, timing and RTC information of muon signals acquired from each layer. Necessary hardware logics are implemented to continuously monitor the health of the detector.

\begin{figure}[htbp]
\centering
\includegraphics[width=.8\textwidth]{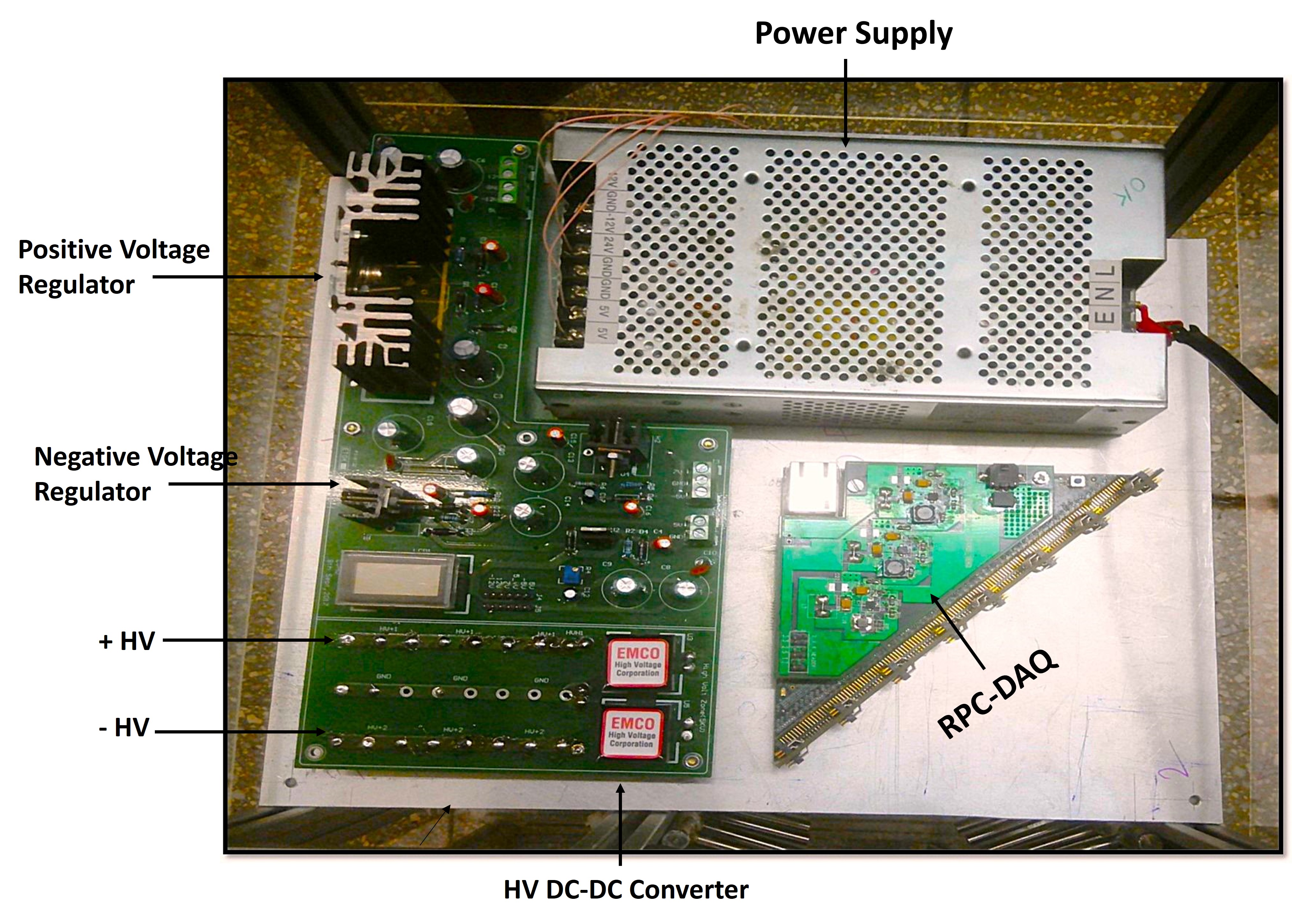}
\caption{Power supply and DAQ electronics mounted on top of the tracker unit.\label{fig:17}}
\end{figure}

Event and monitoring data are transferred to a host PC via Ethernet using a custom server interface implemented in C. The server supports detector configuration (e.g., trigger layer selection, channel masking) and periodically displays health metrics such as strip noise rates and trigger counts. Live event data can also be streamed over the internet for remote monitoring or educational outreach. An integrated Low Voltage and High Voltage (LVHV) distribution module is used to power RPC-DAQ, FE boards and RPCs. 
The DAQ, LVHV and power supply modules are placed on top of the stack as shown in Figure~\ref{fig:17}. 

\section{System Efficiency, Technical Challenges and Student Training}
\label{sec:result}
CMT was evaluated extensively to assess detector performance, data acquisition stability, portability and its utility in student training. The primary metric of interest was the efficiency of the RPCs, determined from muon track reconstruction. Muon tracks were fitted using a least-square method to estimate expected hit positions across each RPC layer. The measured hits were then compared against the expected trajectory to compute efficiency. Table~\ref{tab:1} summarizes the measured efficiency for each RPC layer.

\begin{table}[htbp]
\centering
\caption{Efficiency in \%\label{tab:1}}
\smallskip
\begin{tabular}{|c|c|c|c|c|c|c|c|}
\hline
Layer0 & Layer1 & Layer2 & Layer3 & Layer4 & Layer5 & Layer6 & Layer7 \\
\hline
96.6 & 90 & 48.3 & 90 & 93.2 & 82.5 & 96.5 & 70.6  \\
\hline
\end{tabular}
\end{table}

While two layers exceeded 96\% efficiency, several others showed lower values, ranging between 50--90\%. Notably, some of the lower-efficiency chambers had shown better performance during initial cosmic-ray testing prior to stack assembly. This suggests that performance degradation may have occurred during handling or post-integration. These deviations are due to multiple contributing factors including minor gas leakage leading to degraded avalanche formation and post-assembly mechanical stress. In addition, the large differences in efficiency arose primarily from layer-specific hardware conditions such as electrode contact issues, variation in electronic threshold settings and local noise pickup in certain readout channels.
For long-term operation of the muon telescope, degradation in efficiency and stability is unacceptable. These issues are presently mitigated through periodic maintenance, including inspection and repair of interconnects, along with regular flushing of a fresh gas mixture at least once every five days or whenever a decline in performance is detected. To support this procedure, a pre-mixed 20-liter gas cylinder is permanently assigned to the system and transported with the stack during deployments. Also to ensure all RPCs operate in high and stable efficiency regions for precise muon flux measurements, future revisions of the CMT stack will include improved gas sealing, mechanical reinforcement to minimize stress and contact inconsistencies, and efficiency stability tests conducted before and after each deployment. We will also use high-quality interconnects for handling RPC signals and high voltages, along with proper grounding of the front-end electronics and DAQ systems. Additionally, adopting a more modular stack design will enable in-situ testing of individual RPCs, provide increased space for debugging, and facilitate easier removal and replacement of RPC units.

\begin{figure}[htbp]
\centering
\includegraphics[width=.7\textwidth]{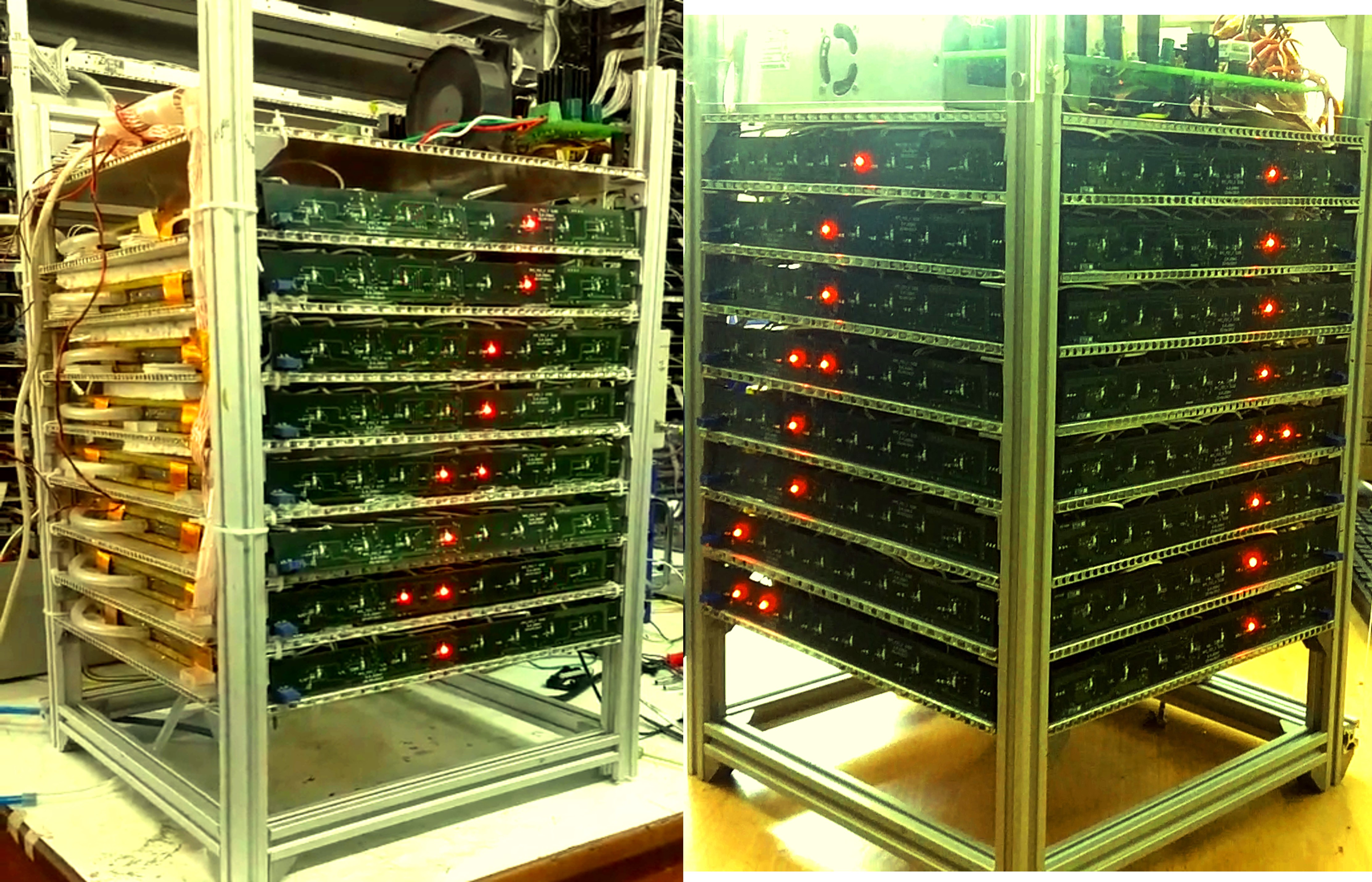}
\caption{LED output on the CMT stack during a live cosmic muon event.\label{fig:18}}
\end{figure}
\begin{figure}[htbp]
\centering
\includegraphics[width=0.9\textwidth]{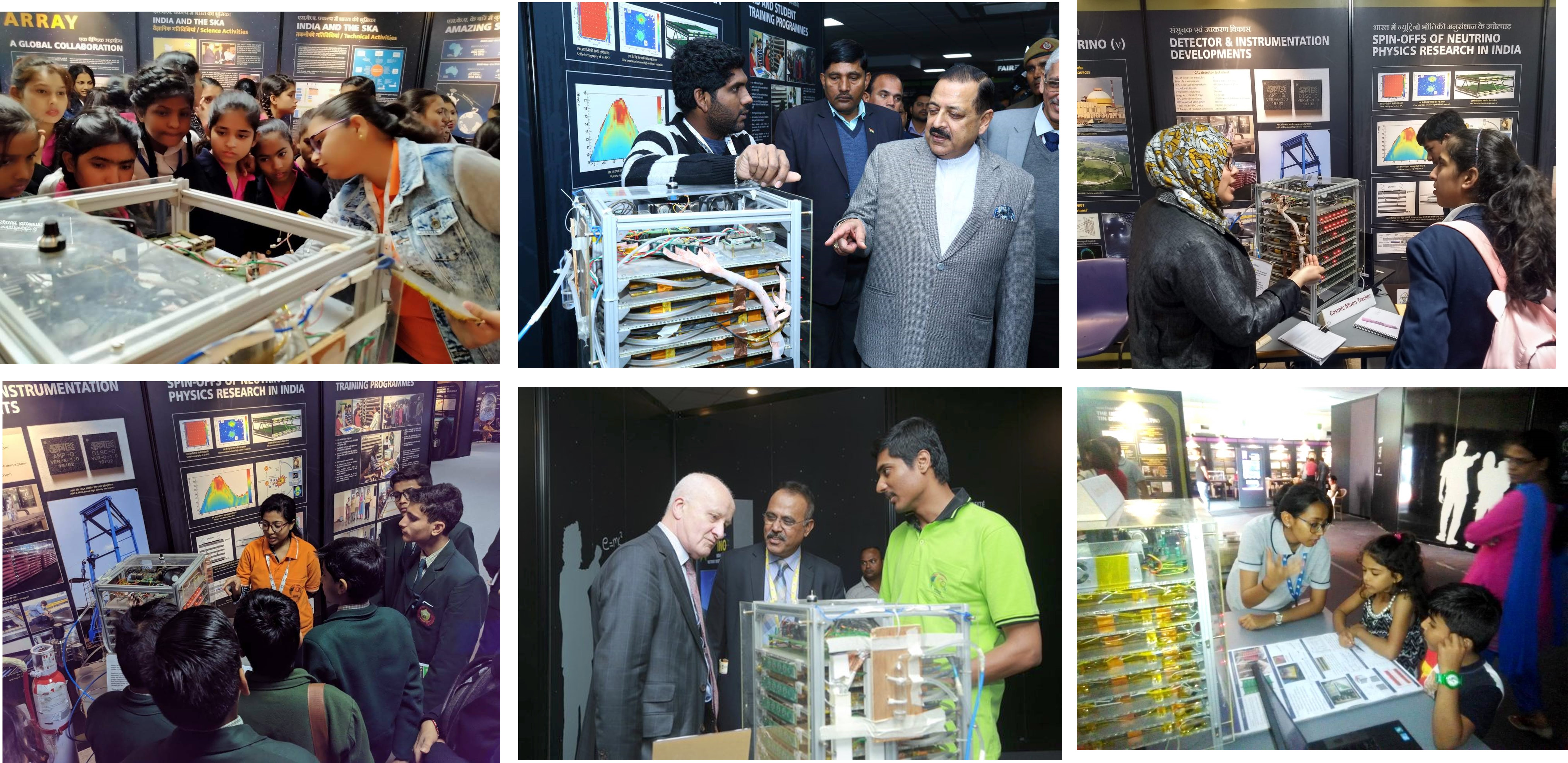}
\caption{Demonstration of the CMT during outreach events, where real-time muon tracking using LED visualization is showcased to students and visitors. \label{fig:19}}
\end{figure}

The LED event display integrated into the DAQ logic provides real-time visual feedback by illuminating LEDs corresponding to muon hits \cite{led_track}, as shown in Figure~\ref{fig:18}. Its stability under varying environmental conditions such as temperature, humidity and pressure was closely monitored. These parameters influence the RPC noise rate and signal multiplicity, thereby impacting the clarity of track visualization. The CMT system was also deployed as a platform for student engagement in hands-on detector development and outreach programs such as Vigyan Samagam as shown in Figure~\ref{fig:19}. 

\section{Conclusion}
\label{sec:con}
CMT is a portable, gas-based detector developed primarily for outreach, educational demonstrations and simple laboratory or field measurements of cosmic ray muons.
The compact and portable design of the CMT makes it ideal for measuring muon flux at different altitudes. With a timing accuracy of $100\, \text{ps}$, it can determine the direction of muon paths and thus can be used for tomography. It can serve as a simple counter for background measurements in mines, etc. The CMT can also function in both vertical and horizontal orientations allowing for comparisons of horizontal and vertical fluxes at different locations as well as any zenith angle. Overall, the CMT offers a reliable and accessible platform for introducing students and the public to particle detection concepts, while also supporting preliminary muon flux measurements and background radiation studies.

\acknowledgments
We extend our gratitude to Antara Prakash and Pratik Barve, intern students, for their valuable contributions to the initial development of the detector and electronics. We also sincerely thank the participants of the RPC lab from the SERB School 2019 held at TIFR for their support in fabricating the RPC detector. Also, we sincerely thank all our present and former INO colleagues especially Puneet Kanwar Kaur, Suraj Kole, Sagar Sonavane, Rajkumar Bharathi for technical support during the design, also we would like to thank members of TIFR, namely K.C. Ravindran, Darshana Gonji, Santosh Chavan and Vishal Asgolkar who supported testing and commissioning. We would like to thank Former INO students Neha Panchal, Abhijit Garai for extending their support during the development. Also, we like to thank former INO directors V.M Datar abd N.K. Mondal for his continuous encouragement and guidance. 



\end{document}